\documentclass[final,5p,times,twocolumn]{elsarticle}

\usepackage{graphicx}

\usepackage{amssymb}

 \biboptions{}

\journal{Physica C: Superconductivity and its Applications}

\begin{document}

\begin{frontmatter}



\title{Thermal treatment of superconductor thin film of the BSCCO system using domestic microwave oven}


\author {J. B. Silveira, C. L. Carvalho, G. B. Torsoni, H. A. Aquino and R. Zadorosny}

\address {Grupo de Desenvolvimento e Aplica\c{c}\~{o}es de Materiais, Departamento de F\'{i}sica e Qu\'{i}mica, Faculdade de Engenharia de Ilha Solteira, Univers Estadual Paulista-UNESP, Caixa Postal 31, 15385-000, Ilha Solteira, SP, Brazil}

\begin{abstract}
 In this work we report the preparation of a superconductor thin film of the BSCCO system using a good quality powder with nominal composition $Bi_{1.8}Pb_{0.4}Sr_2CaCu_2O_x$  which was thermally treated using a domestic microwave oven ($2.45 ~ GHz$, $800 ~ W$). This film was grew on a single crystal of $LaAlO_3$ ($100$) substrate and exhibited a crystalline structure with the c-axis perpendicular to the plane of the substrate. An onset superconducting transition temperature was measured at $80 ~ K$.

\end{abstract}


\end{frontmatter}


\section{Introduction}
\label{Intro}

Since the discovery of high temperature superconductors (HTSs) in 1986 \cite{Bednorz}, a variety of new materials with transition temperatures between $20 ~ K$ and $160 ~ K$, including $La Fe As O_{1-x} F_x$ , $MgB_2$, $La-Ba-Cu-O$ and $Bi-Sr-Ca-Cu-O$ system emerged \cite{Nagamatsu,Tinkhan,Poole,Kamihara,Yang,Luo,Takahashi}. It is well established that the use of precisely controlled thermal treatments is critical to achieve the desired phase, as well as the structural, electrical and magnetic properties. Studies demonstrate that ceramic materials thermally processed in microwave oven can decrease the processing time when compared with heat treatment in conventional furnaces \cite{Santos,Rao}.

Microwaves are electromagnetic waves with frequencies in the range of $0.300 ~ GHz$ and $300 ~ GHz$. They are used in a variety of technological areas including civil and military radars, mobile phones, microwave ovens and so on \cite{Rao}. The first use of microwave ovens for non domestic applications date to the 1960's when it was used for drying red ceramics. Analytical chemists used microwaves in laboratorial research and in preparation of samples in the 1970's \cite{Keyson}. Microwave processing has received special interest for synthesis and thermal processing of organic and inorganic compounds to take advantage of energy savings, reduced processing times and improvement of microstructure of the compounds \cite{Santos,Rao,Keyson,Saita,dosSantos,Menezes,Vasconcelos}.

The heating process with microwave involves the absorption of radiation by depolarization of the polar molecules and thermal convection. The absorption results from the interaction between the electromagnetic wave and the molecule dipole moment \cite{Barbosa,Zlotorzynski,Simoes}. The microwave electric field also induces an alignment between the molecule and the electric field. The phenomenon of dielectric relaxation of the molecule, what means the molecule oscillates with the field, dissipates energy by Joule effect, as a consequence, large amounts of energy can be absorbed and transformed in heat \cite{Barbosa,Zlotorzynski}.

Sources with frequency of $2.45 ~ GHz$ are readily available because they are typically used in domestic microwave ovens to heat the water molecule at its resonance frequency. High efficiency can be reached in this process if the samples were constituted by molecules with relaxation frequency similar as the wave frequency of the microwave radiation. Some works have focused on the differences between materials synthesized or heat treated by microwave oven and by conventional furnaces. Santos \textit{et al.} \cite{Santos} and Sim\~{o}es \textit{et al.} \cite{Simoes,Simoes2} observed significant improvement in the structural properties of materials prepared in microwaves ovens.

In the present work we report the preparation of superconductor thin film using a good quality powder of BSCCO system with stoichiometry $2212$, grown on a $La Al O_3$ crystalline substrate and using a domestic microwave oven as heater source. Microstructure, chemical composition, electric and magnetic properties were investigated.

\section{Experimental Procedure}
\label{Exp}

\begin{figure}
\includegraphics[width=1\columnwidth,height=0.5\linewidth]{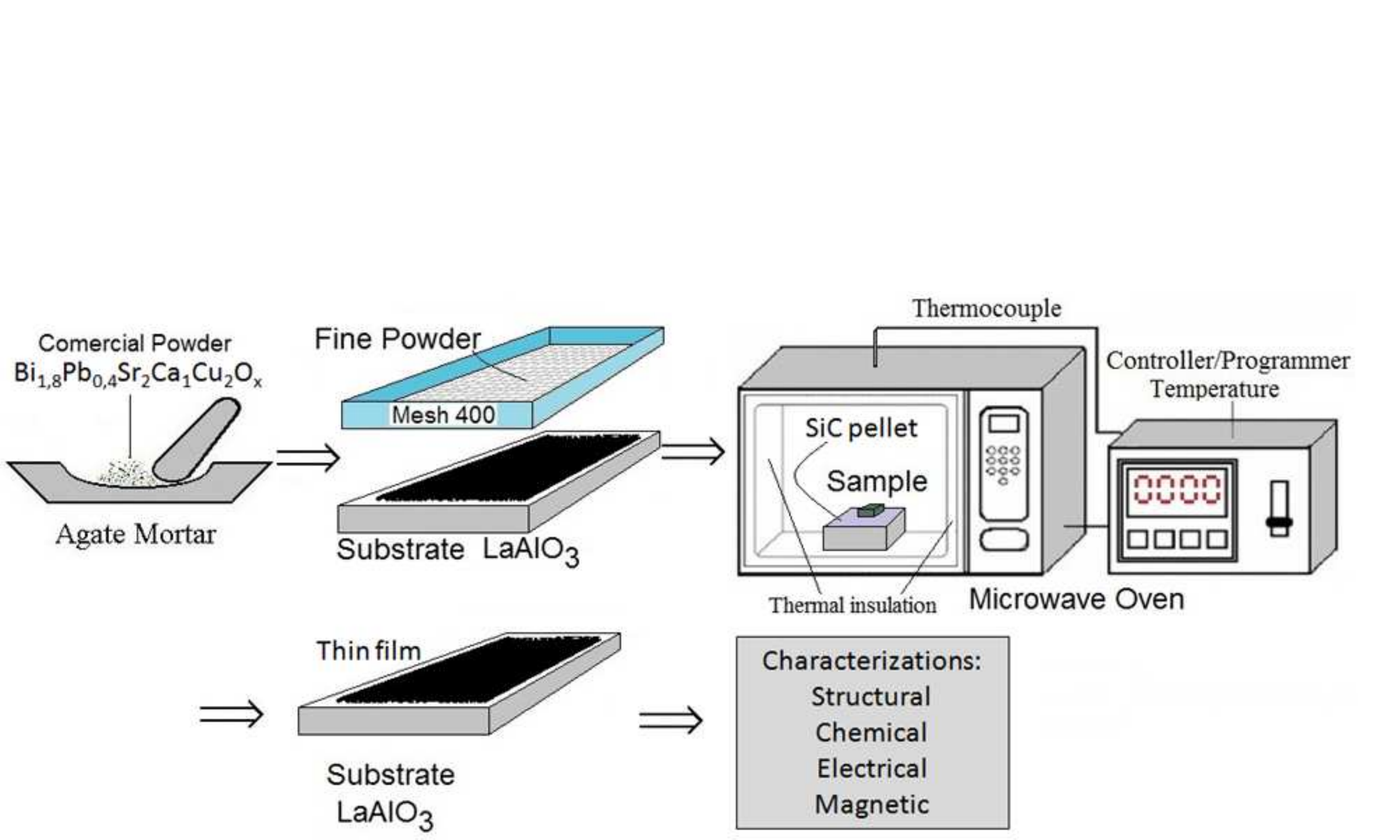}
\caption{Experimental scheme for the preparation of a superconductor thin film using a domestic microwave oven.\label{fig:1}}
\end{figure}

Our study was focused on the effect of the thermal treatment using a domestic microwave oven on commercial ceramic powder of the superconductor BSCCO system denominated $2212$ and used to obtain a thin film which presented a critical temperature $T_c \approx 80 ~ K$.  To produce such sample, $5 ~ mg$ of commercial powder with nominal composition $Bi_{1.8} Pb_{0.4} Sr_2 Ca Cu_2 O_x$, fabricated by Superconductive Components Inc. ($Lot\# SCI3130SCP3$) was spread on a $La Al O_3$ crystalline substrate with orientation ($100$). The powder distribution was controlled using a $400$ mesh sieve (aperture of $37\mu m$). The Fig.\ref{fig:1} shows a scheme of the film preparation. The film was thermally treated at $800^{o}C$ for $10$ minutes using a domestic microwave oven (Panasonic model $NN-S46BK$, $800 ~ W$ and $2.45 ~ GHz$) at $20^{o}C$ per minute heating/cooling rate. A small pellet of $SiC$ ($30$x$25$x$5 ~ mm^3$) was used as a susceptor \cite{Simoes2} in order to absorb the microwave energy and rapidly transfer the heat to the film. The microwave oven was lined with a thermal blanket of alumina to protect the metallic inner structure of high temperatures of the order of $800^{o}C$. A K-type thermocouple was placed in contact with the susceptor to control the temperature in the sample.

The structural characterization was performed by X-ray diffraction (XRD) in a Shimadzu diffractometer model $XRD-6000$ with copper filter $K_\alpha$ and step of $0.02^o$. The microstructural analysis was performed by scanning electron microscopy (SEM) in a $SiLi$ detector with resolution of $68 ~ eV$ and ZAF quantitative method was used in the energy dispersive spectroscopy (EDS) compositional analysis.

To measure the resistance vs. temperature, $R(T)$, it was used the DC four probe method with an applied transport current of $10 ~ mA$, using a programmable voltage/current source model $228A$ from Keithley Instruments.
AC-susceptibility vs. temperature, $\chi(T)$, with different excitation fields, $H_{AC}$, and magnetization (normalized by the applied field, $H_{DC}$) vs. temperature for different $H_{DC}$ were performed in a Quantum Design magnetometer, PPMS model $6000$.

\begin{figure}
\includegraphics[width=0.9\columnwidth,height=0.8\linewidth]{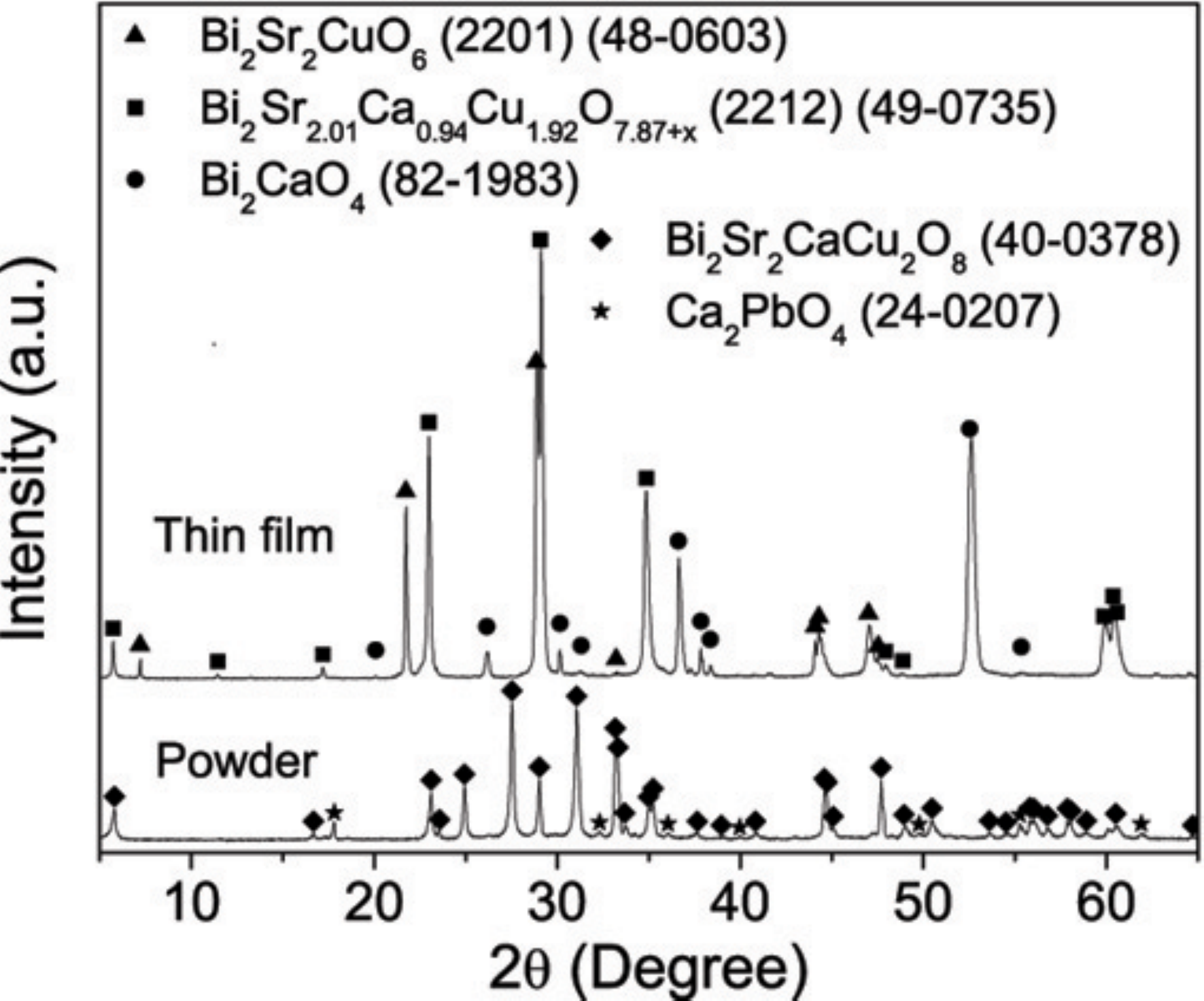}
\caption{X-ray patterns with indexed crystallographic peaks for  the BSCCO commercial powder and for the film synthesized by microwaves. Notice that in this diffractogram various peaks shown orientation $(00l)$, such as in $2\theta$ $= 5.7^o$ $(002)$; $23.0^o$ $(008)$; $28.9^o$ $(0010)$; $34.8^o$ $(0012)$; $47.0^o$ $(0016)$; $59.8^o$ $(0020)$ which indicates a preferential grown epitaxially with the $La Al O_3$ $(100)$ substrate \label{fig:2}.}
\end{figure}

\section{Results and Discussion}
\label{Resul}

Fig.\ref{fig:2} shows the XRD patterns for the powder and for the film. The diffractogram confirms the 2212 phase in the powder, which presents also a small amount of the secondary phase $Ca_2 PbO_4$ \cite{Hosseini,Wong}. In the diffractogram of the film, it is possible to identify several crystallographic peaks of the $2212$ phase from which many of them preferentially oriented relative to the $(001)$ plane, i.e., the crystals grown epitaxially with the substrate as expected. Some peaks indicate the presence of the $2201$ phase that is due to the range of temperature treatment for which the sample was submitted. Moreover, also occurs the presence of a small portion of the secondary phase such as $Bi_2CaO_4$ \cite{Wong,Majewski,Polasek,Torrance,Idink,Natalisora,Tromel}. A calculation using the Debye-Scherrer equation \cite{Cullity} shown that the typical crystallite size was $81 ~ nm$ and $54 ~ nm$ for the film and powder respectively. This implies that the thermal treatment with microwaves induce the coalescence between the crystallites.

\begin{figure}
\includegraphics[width=1\columnwidth,height=1\linewidth]{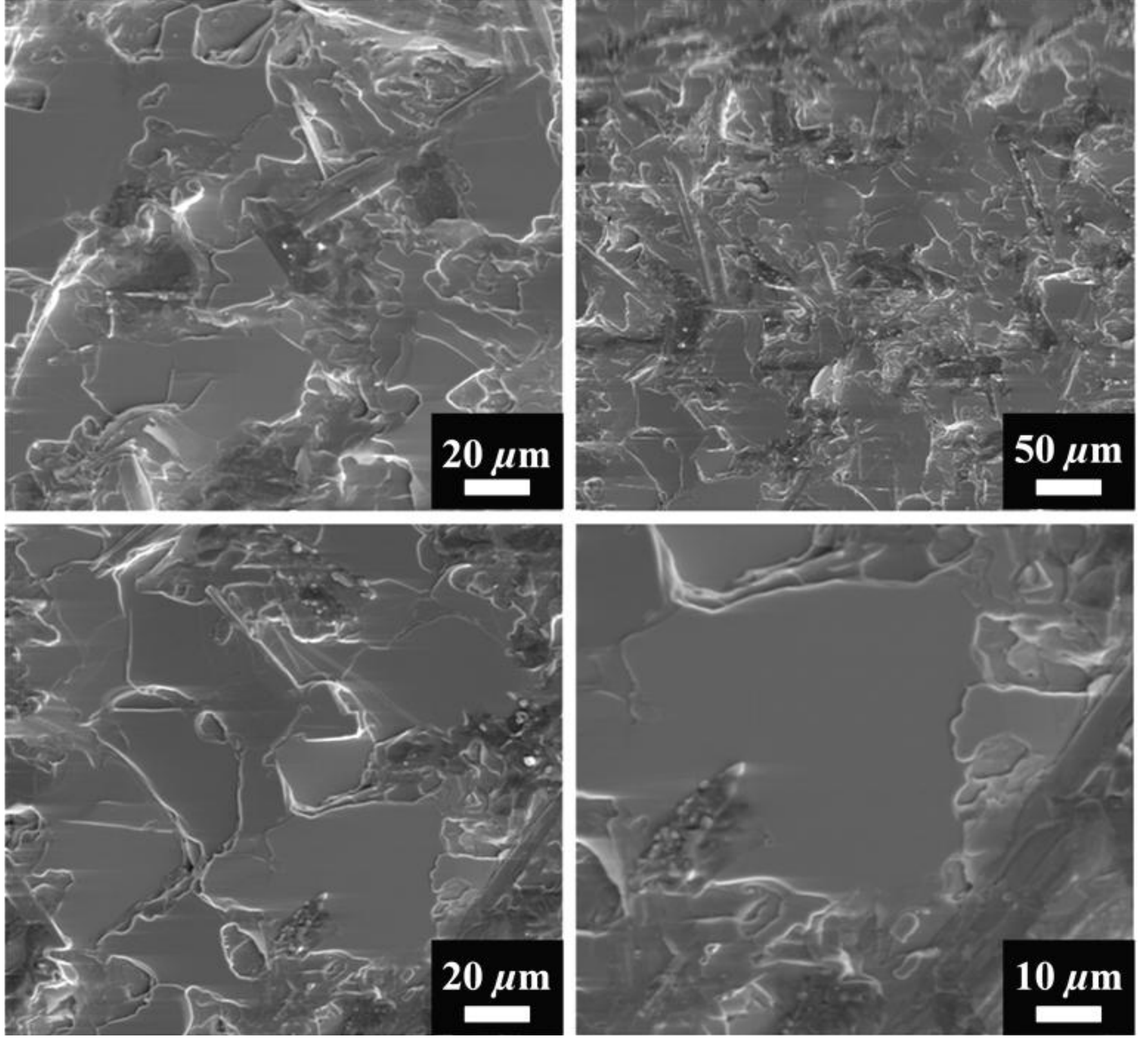}
\caption{SEM micrographs of different regions of the sample. Up panels (from left to right side) the magnifications are 500x and 200x respectively. The left and down panel shows an image amplified by 200x and the right-down panel shows a zoom (amplification of 1000x) on the same region focusing on the plate-like grain of dimensions around $50$x$50 ~ \mu m^2$.}
\label{fig:3}       
\end{figure}

Micrographs using SEM images at different magnifications illustrate the morphology of the BSCCO film, as shown in Fig.\ref{fig:3}. The presence of large plate-like grains with layered growth typical of (Bi, Pb)-$2212$ phase of dimensions around $50$x$50 ~ \mu m^2$ and porous regions are observed \cite{Majewski,Vinu}. Such behavior indicate that the thermal treatment using microwave induce the connectivity between the crystallites forming the large grains observed. Notice that in this study we are considering the classification used by Cullity, i.e., crystallites are little single crystals and grains are formed by the agglomerate of those crystallites \cite{Cullity}. This microstructure is not anticipated that such sample could not carry large current densities as a result of a wide distribution of grains size and the inhomogeneity of intergrain coupling (weak-links, WLs).

\begin{table}
\caption{Data obtained after analyzing a film area of $450$x$450 ~ \mu m$ with $200$x magnification.}
\label{tab:1}       
\begin{tabular}{ccccc}
\hline\noalign{\smallskip}
Element&Mass&Atomic&Nominal&EDS\\
\noalign{\smallskip}\hline\noalign{\smallskip}
Bi M&49.79&25.75&1.80&1.85\\
Pb M&5.68&2.96&0.40&0.21\\
Sr L&21.65&26.71&2.00&1.92\\
Ca K&5.67&15.29&1.00&1.10\\
Cu K&17.21&29.28&2.00&2.11\\
Total&100.00&99.99&7.20&7.20\\
\noalign{\smallskip}\hline
\end{tabular}
\end{table}

\begin{table}
\caption{Data obtained after analyzing a film area of $90$x$90 ~ \mu m$ with $1000$x magnification.}
\label{tab:2}       
\begin{tabular}{ccccc}
\hline\noalign{\smallskip}
Element	& Mass	& Atomic	& Nominal	& EDS  \\
\noalign{\smallskip}\hline\noalign{\smallskip}
Bi M&	61.51& 38.34&	1.80&	2.76 \\
Pb M&	6.33&	3.98&	0.40&	0.29\\
Sr L&	19.37&	28.79&	2.00&	2.07\\
Ca K&	2.21&	7.18&	1.00&	0.52\\
Cu K&	10.59	&21.7&	2.00&	1.56\\
Total&	100.01&	99.99&	7.20&	7.20\\
\noalign{\smallskip}\hline
\end{tabular}
\end{table}

EDS was used to determine the nominal composition of the same sample whose microstructure was described earlier. Tables \ref{tab:1} and \ref{tab:2} show the probably composition for two areas of the sample, $450$x$450 ~ \mu m^2$ and $90$x$90$ $\mu m^2$ respectively, analyzed using three iterations of the measurement sequence, corresponding to $Bi_{1.85} Pb_{0.21} Sr_{1.92} Ca_{1.10} Cu_{2.11} O_x$ and $Bi_{2.76} Pb_{0.29} Sr_{2.07} Ca_{0.52} Cu_{1.56} O_x$ respectively.

\begin{figure}
\includegraphics[width=1\columnwidth,height=0.7\linewidth]{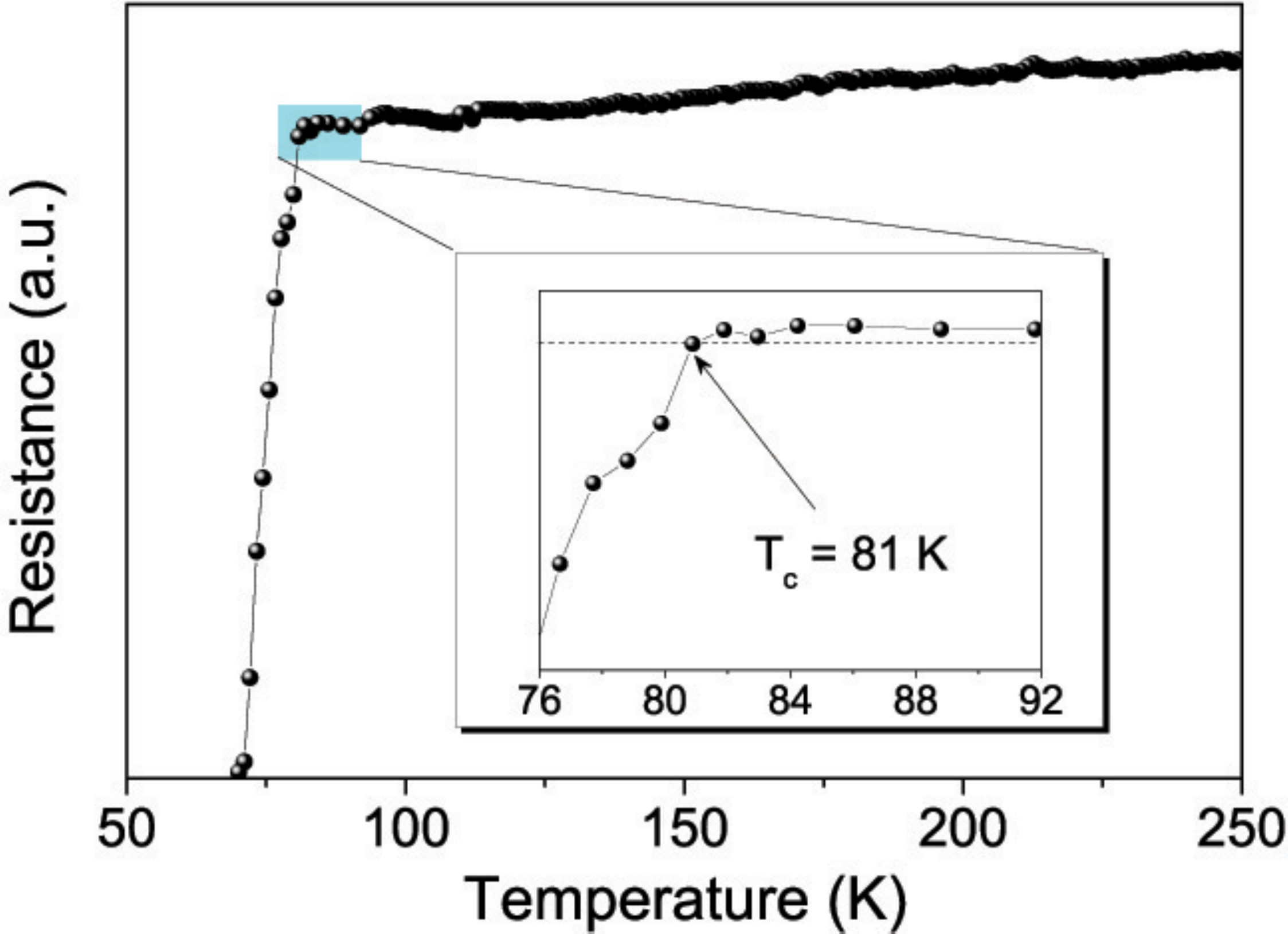}
\caption{Resistence as a function of the temperature for the BSCCO film heat. In the normal state the resistance presents an ohmic behavior. The superconducting transition starts at $T_c=81 ~ K$.}
\label{fig:4}       
\end{figure}

\begin{figure}
\includegraphics[width=1\columnwidth,height=0.7\linewidth]{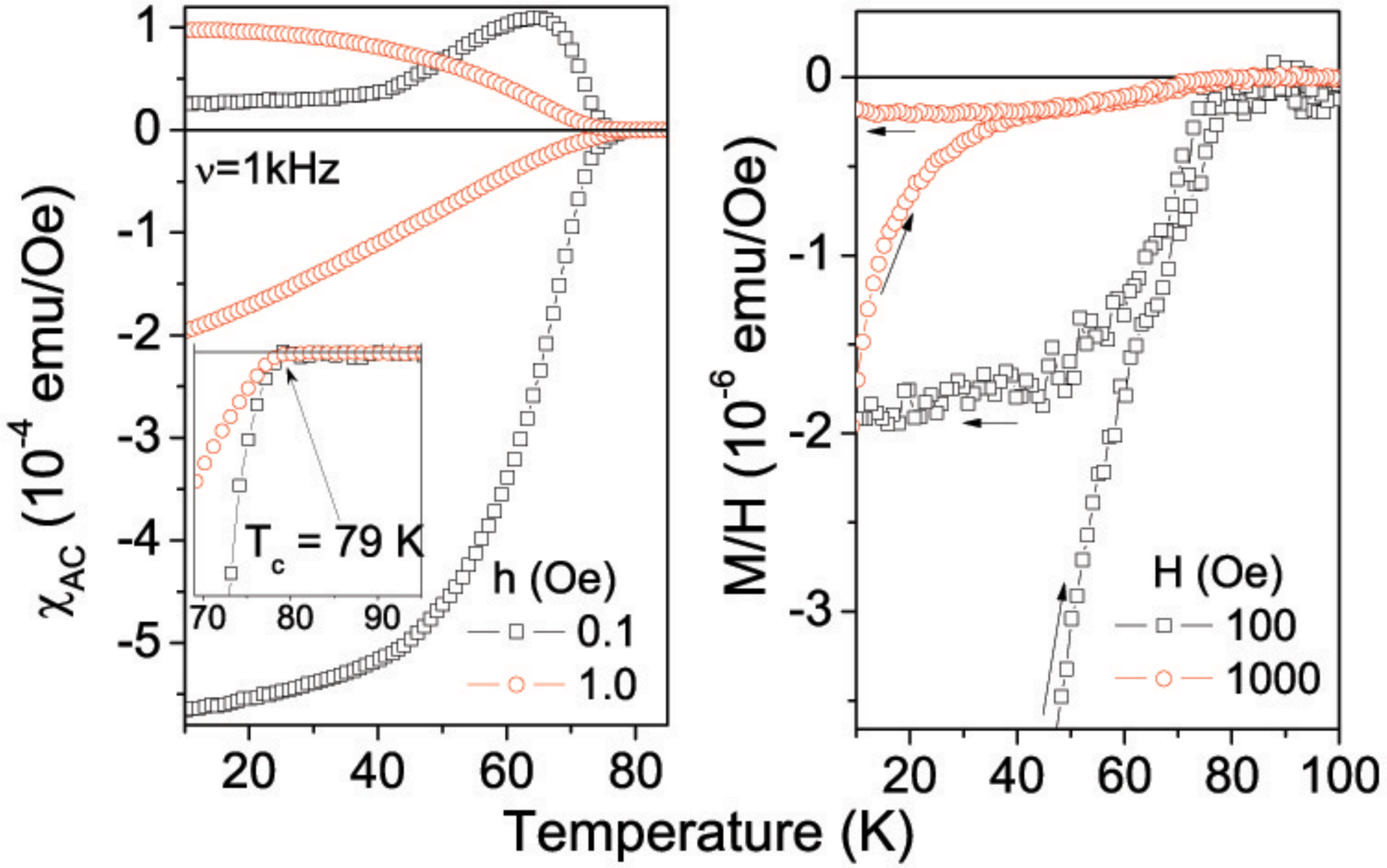}
\caption{Temperature dependence of the AC-susceptibility and the magnetization (normalized by $H$). The significant dependence of such curves with the applied and excitation fields are related to inhomogeneity of the WLs.}
\label{fig:5}       
\end{figure}

The superconducting properties of the film were studied by electrical and magnetic measurements. Fig.\ref{fig:4} shows a typical $R(T)$ response. In the normal state the resistance presents an ohmic behavior between $250 ~ K$ (higher temperature measured) and $T_c = 81 ~ K$, from which a superconducting transition takes place. This value for $T_c$ is less than that reported to $2212$ phase ($T_c = 85 ~ K$) \cite {Cyrot}, which also implies in a broad superconducting transition ($\Delta T_c$) as we can see in Fig.\ref{fig:4}.

To measure the magnetic properties of the sample, a slice of $3$x$3$ $mm^2$ was used. In Fig.\ref{fig:5}(a) are shown curves of AC-susceptibility, $\chi(T)= \chi´(T)+ i \chi´´(T)$, obtained for $H_{AC} = 0.1 ~ Oe$ and $H_{AC} = 1.0 ~ Oe$ with frequency of $\nu =1 ~ kHz$. One can see a dependence of $\chi(T)$ with $H_{AC}$. This dependence is characteristic of granular samples due to the WLs. However, the significant difference between those curves is related to the inhomogeneity of the WLs caused, e.g., by non superconducting materials formed in the sintering  process as well as the wide distribution of grains size (see the micrographs of Fig.\ref{fig:3}). Although, there is no change in $T_c$ which implies that the intragranular current density induced by $H_{AC}$ is sufficient to shield the grains.

Such behavior is also present in magnetization (normalized by $H_{DC}$) vs. temperature curves for different $H_{DC}$, as shown in Fig.\ref{fig:5} (b). Since WLs limit the shielding and/or the transport current and this limitation is strongly dependent of $H_{DC}$ even cooled down the film up to $10~K$, only a single superconducting-normal estate transition takes place. In our case, this means that for $H_{DC} = 100 ~ Oe$ and $1 ~kOe$ there is no intergranular current (WLs are not working) and the response is due to the intragranular portion. The irreversibility between ZFC (zero field cooling) and FC (field cooling) is characteristic of ceramic samples. However, the reversible behavior is associated with the fusion of the vortex lattice \cite{Silva,Rodrigues,Jung,Mohamed,Sutjahja}  which, in our case, probably occurs into the plates.

\section{Conclusions}
\label{Conclu}

 We prepared a BSCCO superconductor thin film from commercial powder which was thermally treated using a domestic microwave oven. Large plate-like grains with dimensions around $50$x$50~ \mu m^2$ and porous regions were observed in SEM micrographs. The resulting sample presented a $T_c$ of the order of $80 ~ K$ which was obtained by electrical and magnetic techniques. This temperature and the broad superconducting transition are due to the presence of the $2201$ phase as observed in the XRD measurements. As indicated by the SEM images, the magnetic measurements confirm the deficient connectivity between the plates (weak-links) which is close linked to the capability of the material in transport current, i.e., better connectivity-higher transport currents. We can also conclude that microwave thermal treatment can be performed significantly faster than that used in conventional furnaces, which implies in a save of electrical energy

\section{Acknowledgments}
\label{Acknow}

This work was supported by FAPESP, Fundunesp, PROPe, CNPq and Capes. We are grateful to Polymer Group of Physics and Chemistry Department by the XRD measurements; Prof. Dr. Sebasti\~{a}o Ribeiro from EEL-USP by the microwave susceptors and Prof. Dr. Wilson Aires Ortiz, leader of "Superconductivity and Magnetism Group" at UFSCar by magnetic measurements and discussions.



\bibliography{<your-bib-database>}



\end{document}